\documentclass[prc,aps,floatfix,twocolumn,superscriptaddress,nofootinbib,showpacs,preprintnumbers,
amsmath,amssymb,amsfonts,widetable] {revtex4-2}
\usepackage{bm}
\usepackage{soul}
\usepackage{array}
\usepackage{lipsum}
\usepackage{relsize}
\usepackage{mathrsfs}
\usepackage{amssymb}
\usepackage{amsmath}
\usepackage{graphicx}
\usepackage[usenames]{color}
\usepackage{pslatex}
\usepackage{booktabs}
\usepackage{physics}
\usepackage{float}
\usepackage{tcolorbox}
\usepackage{ulem}

\newcommand{\fF}{\mathlarger{f}_{\raisebox{-0.5pt}{\tiny F}}}
\newcommand{\fS}{\mathlarger{f}_{\raisebox{-0.5pt}{\!\tiny SF}}}
\newcommand{\FF}[1]{\mathlarger{F}_{\raisebox{-0.5pt}{\!\tiny #1}}}

\begin{document}

\title{The Matter Radius of \textsuperscript{132}Sn and the CREX--PREX Dilemma}
\author{J. Piekarewicz}
\affiliation{Department of Physics, Florida State University, Tallahassee, FL 32306, USA}

\date{\today}

\begin{abstract}
The density dependence of the nuclear symmetry energy remains a central open problem 
in nuclear physics. Parity violating electron scattering experiments have provided largely 
model-independent determinations of the neutron skin thickness of both ${}^{48}$Ca and 
${}^{208}$Pb, whose consistent theoretical interpretation remains challenging. A new 
measurement of the matter radius of the unstable, doubly magic nucleus ${}^{132}$Sn 
provides an important additional constraint. Using a representative set of covariant energy 
density functionals spanning a wide range of isovector properties, we show that 
at least one of these models can simultaneously reproduce the charge and matter radii of
${}^{132}$Sn. When interpreted together with the PREX and CREX results, the new 
measurement---much like CREX---favors a relatively soft symmetry energy. These findings 
underscore the need for an independent confirmation of the PREX result, such as the 
anticipated MREX campaign at the MESA facility.
\end{abstract}

\maketitle


The density dependence of the nuclear symmetry energy remains one of the
central open questions in nuclear science, with profound implications for both 
neutron-rich nuclei and neutron stars; see Refs.\cite{Steiner:2004fi,Tsang:2012se,
Horowitz:2014bja,BaldoBurgio2016,Li:2021} and references contained therein.

Specifically, parity violating electron scattering experiments on both $^{208}$Pb 
(PREX)\,\cite{Abrahamyan:2012gp,Horowitz:2012tj,Adhikari:2021phr}  and $^{48}$Ca 
(CREX)\,\cite{Adhikari:2022kgg} at the Thomas Jefferson National Accelerator Facility 
have provided the first model-independent determinations of the weak form factor at 
precisely selected momentum transfers. Parity-violating scattering is uniquely powerful 
because it avoids the strong-interaction uncertainties inherent in hadronic measurements. 
While the transition from the weak form factor to the neutron radius---and the resulting 
neutron skin thickness---is mildly model dependent\,\cite{Adhikari:2021phr}, these 
measurements currently offer the cleanest determination of the neutron radii of 
$^{48}$Ca and $^{208}$Pb. 

Unfortunately, a combined interpretation of these results has proven to be remarkably 
challenging. While PREX favors a relatively thick neutron skin in $^{208}$Pb that
is consistent with a stiff symmetry energy\,\cite{Adhikari:2021phr,Reed:2021nqk}, 
CREX measured a significantly thinner skin in $^{48}$Ca\,\cite{Adhikari:2022kgg}, 
suggesting a softer symmetry energy near saturation density. Reconciling these 
seemingly contradictory findings within a unified theoretical framework has exposed 
significant tensions in both density functional theory and ab initio 
approaches\,\cite{Hu:2021trw,Reinhard:2022inh,
Zhang:2022bni,Mondal:2022cva,Papakonstantinou:2022gkt,Yuksel:2022umn,Li:2022okx,
Thakur:2022dxb,Miyatsu:2023lki,Reed:2023cap,Sammarruca:2023mxp,Salinas:2023qic,
Yue:2024srj,Zhao:2024gjz,Roca-Maza:2025vnr,Kunjipurayil:2025xss}.

In this context, the recent experimental extraction of the matter radius of the doubly magic 
nucleus $^{132}$Sn provides a critical new benchmark\,\cite{Hijikata:2026xwa}. For decades
$^{132}$Sn---with closed shells of 50 protons and 82 neutrons, and a lifetime of nearly 40 
seconds---has been a cornerstone of nuclear physics due to its robust ``doubly magic'' 
character\,\cite{Gorges:2019wzy,Gustafsson:2025xpd}. Because of its structural simplicity, 
$^{132}$Sn has long served as an ideal testing ground for theoretical 
models\,\cite{Arthuis:2020toz,Hu:2021trw} and the investigation of exotic collective modes
sensitive to the nuclear symmetry energy\,\cite{Adrich:2005,Piekarewicz:2006ip,
Klimkiewicz:2007zz,Paar:2007bk,Piekarewicz:2010fa,Carbone:2010az,Savran:2013bha}

Detailed structural information on $^{132}$Sn began to emerge in 2005 with a precise 
measurement of its charge radius. Laser spectroscopy studies of neutron-rich tin isotopes 
performed at the ISOLDE facility yielded a precisely measured charge radius of 
$R_{\rm ch}^{132}\!=\!4.709(7)\,{\rm fm}$\,\cite{LeBlanc:2005ik}. Soon thereafter, reactions 
on nuclei neighboring $^{132}$Sn provided direct access to single-particle states outside the 
$N\!=\!82$ shell closure\,\cite{Jones:2010ci,Vaquero:2020mlx}. These experiments showed 
that the spectroscopic strength associated with the valence neutron orbitals is consistent with 
a pronounced shell closure, thereby reinforcing the magic nature of $^{132}$Sn.

Beyond nuclear structure, $^{132}$Sn also plays an important role in nucleosynthesis, as it
lies close to the path of the r-process responsible for producing roughly half of the heavy 
elements\,\cite{Jones:2010ci,Jones:2011kp}. Taken together, these experimental advances 
have strongly influenced theoretical developments. A unified description of the single-particle 
spectrum, the charge radius, and now the matter radius is essential for developing a comprehensive 
picture of how neutrons and protons are distributed within the short-lived, neutron-rich nucleus 
$^{132}$Sn.

With an isospin asymmetry even larger than that of $^{208}$Pb, ${}^{132}$Sn offers enhanced 
sensitivity to isovector dynamics while remaining amenable to both density functional theory 
and ab initio calculations\,\cite{Arthuis:2020toz,Hu:2021trw}. The measured matter radius, 
$R_{\rm m}^{132}\!=\!4.758^{+0.023}_{-0.024}\,{\rm fm}$, when combined with the charge radius 
points---much as in the case of $^{48}$Ca---to a relatively thin neutron skin. As relevant, Hijikata 
and collaborators concluded that ``there are no theoretical calculations consistent with both matter 
and charge radii within the experimental errors"\,\cite{Hijikata:2026xwa}. Hence, any realistic 
description of the nuclear symmetry energy must now account not only for the neutron skins 
of $^{48}$Ca and $^{208}$Pb, but also for the correlated charge-matter radius systematics 
emerging in $^{132}$Sn. The delicate interplay among these nuclei therefore provides a uniquely 
stringent test of modern energy density functionals and ab initio approaches. The combined 
constraints from $^{48}$Ca, $^{132}$Sn, and $^{208}$Pb thus define a powerful triplet of 
doubly-magic nuclei that sharpens our understanding of isovector dynamics and of the density 
dependence of the symmetry energy.


The results presented in this letter were obtained within a covariant framework based on a 
Lagrangian density written in terms of nucleon, photon, and three meson fields that mediate the 
nuclear interaction. The structure of the Lagrangian density has been documented extensively in 
earlier publications\,\cite{Horowitz:2000xj}, while details of the specific models employed in this 
work may be found in Refs.\cite{Fattoyev:2010mx,Chen:2014sca,Chen:2014mza}.

\begin{center}
\begin{figure}[h]
\centering
\bigskip
\includegraphics[width=0.49\textwidth]{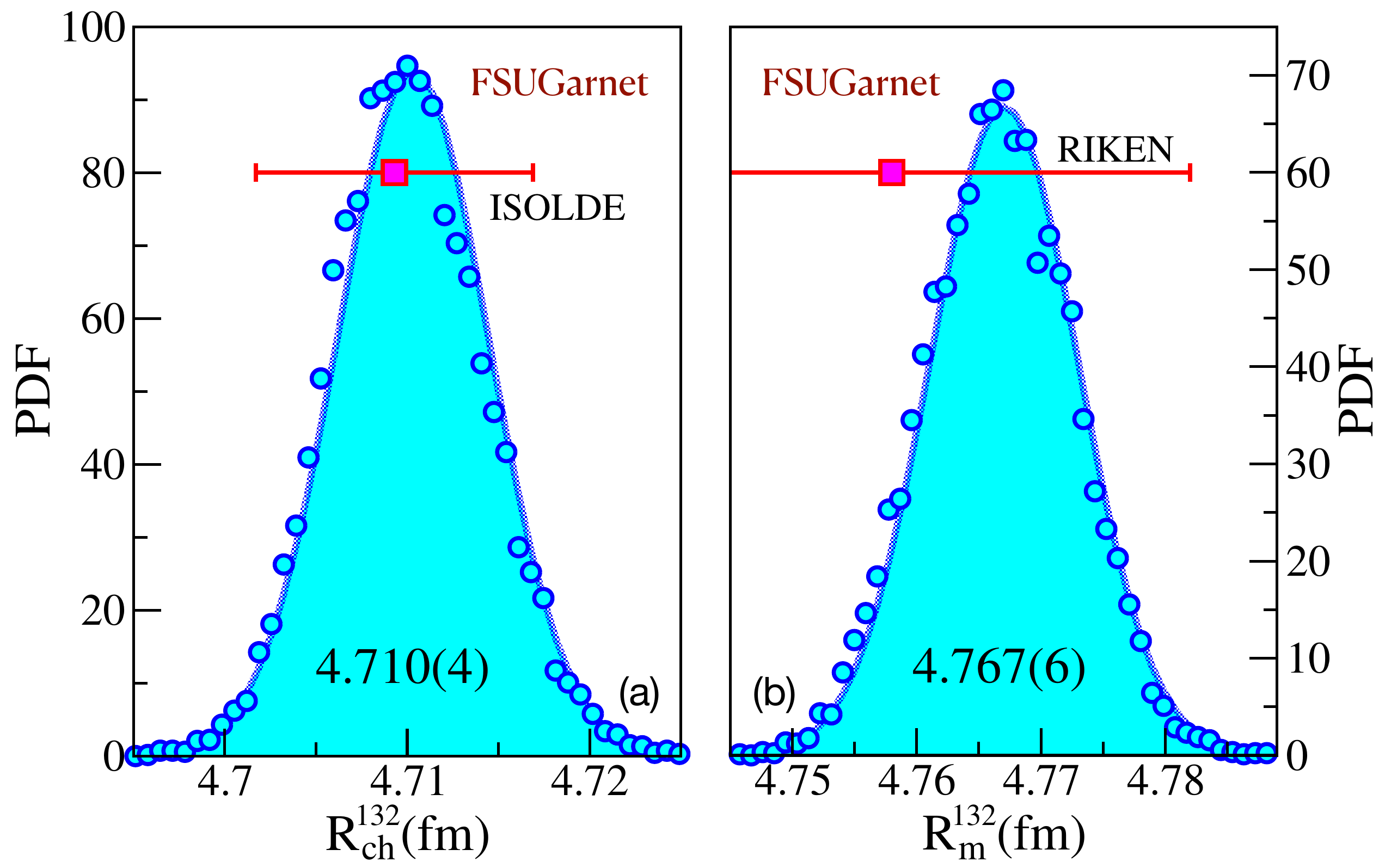}
\caption{Theoretical predictions for the charge and matter radii of ${}^{132}$Sn.
	      The probability distribution functions were generated with 
	      FSUGarnet\,\cite{Chen:2014mza} and the experimental data are from 
	      ISOLDE\,\cite{LeBlanc:2005ik} and RIKEN\,\cite{Hijikata:2026xwa}, 
	      respectively.}
\label{Fig1}
\end{figure}
\end{center}

In Fig.\ref{Fig1} we display predictions for the probability distribution functions of the charge and 
matter radii of ${}^{132}$Sn as obtained with FSUGarnet---a covariant energy density functional 
introduced in Ref.\cite{Chen:2014mza}. The distributions were generated from 10,000 configurations 
sampled according to the covariance matrix obtained from the optimization of the functional. The 
circles represent the histogram, while the solid curve denotes the associated Gaussian fit. Also 
shown are the experimental values for the charge\,\cite{LeBlanc:2005ik} and matter 
radii\,\cite{Hijikata:2026xwa}. Contrary to the assertion by Hijikata \textit{et al.}, theoretical calculations 
exist that are consistent with both the measured charge and matter radii. As we elaborate below, the 
real challenge for the theoretical community is to reconcile all the existing measurements, namely, 
PREX, CREX and now the RIKEN results.

Together with the charge and matter radii displayed in Fig.\ref{Fig1}, we now list in Table \ref{Table1} the 
charge, matter, neutron, and neutron skin radii of $^{132}$Sn as predicted by three representative covariant 
energy density functionals: FSUGarnet, RMF022, and FSUGold2\,\cite{Chen:2014mza, Chen:2014sca}. That 
all three models agree on their predictions for the charge radii is not surprising, given that the charge radius of 
$^{132}$Sn was incorporated into the calibration of the functionals. Note that the charge radius was defined in 
the calibration of the various functionals simply as $R_{\rm ch}^{2}\!=\!R_{p}^{2}+r_{\rm ch}^{2}$, where $R_{p}$ 
is the point proton radius and $r_{\rm ch}\!=\!0.88\,{\rm fm}$ is the single-proton charge radius---an adopted
value that is slightly larger than the current CODATA value of $r_{\rm ch}\!=\!0.84\,{\rm fm}$. The models
primarily disagree in the calibration of the isovector sector: FSUGarnet predicts the softest symmetry energy 
and the thinnest neutron skin, while in turn, FSUGold2 predicts the stiffest symmetry energy and the thickest
neutron skin. Note that a ``soft" or ``stiff" symmetry energy refers to the slope parameter of the symmetry 
energy at saturation density ($L$), a quantity that correlates strongly with the value of the neutron 
skin thickness of heavy nuclei\,\cite{Brown:2000,Furnstahl:2001un,RocaMaza:2011pm}. The slope of the
symmetry energy for the functionals selected in this work are: $L\!=\!50.96, 63.52, 112.84\,{\rm MeV}$, for 
FSUGarnet, RMF022, and FSUGold2, respectively.

\begin{center}
\begin{table}[h]
\begin{tabular}{|l|c|c|c|c|}
\hline\rule{0pt}{2.5ex}   
\!\!Model   &  $R_{\rm ch}(\rm fm)$  &  $R_{\rm m}(\rm fm)$  & 
                     $R_{\rm n}(\rm fm)$ &  $R_{\rm skin}(\rm fm)$  \\
\hline
\hline
FSUGarnet &  4.710(4)   &   4.767(6)   & 4.850(10)   &  0.223(12) \\
RMF022     &   4.706(4)  &   4.799(7)   &  4.904(11)  &  0.280(13)  \\
FSUGold2  &  4.707(5)   &   4.846(13) & 4.976(19)  &  0.352(16)   \\
\hline
Experiment &  4.709(7)   & 4.758(24)  & 4.824(38) & 0.178(39) \\
\hline
\end{tabular}
\caption{Predictions for the charge, matter, neutron, and neutron‑skin radii 
             of ${}^{132}$Sn obtained with the three covariant energy density 
             functionals employed in this work. The uncertainties in the experimental 
             neutron radius and neutron skin thickness were obtained via standard 
             error propagation.}
 \label{Table1}
\end{table}
\end{center}
Although charge radii are fundamental observables for the calibration of energy density functionals, 
a more stringent test of the theoretical framework is provided by the corresponding densities and 
form factors. To place the discussion in the proper context, we highlight several useful properties of 
the ``symmetrized'' two-parameter Fermi function. Although it is practically indistinguishable from the 
standard two-parameter Fermi (2pF) form when the half-density radius $c$ is much larger than the 
surface diffuseness $a$, the symmetrized version exhibits significantly improved analytic 
properties\,\cite{Sprung:1997}. Starting from the conventional 2pF distribution, the symmetrized 
form may be written as follows:
\begin{equation}
 \fS(r) \equiv \fF(r) + \Big[\fF(-r)\!-\!1\Big] =
 \frac{\sinh(c/a)}{\cosh(r/a)+\cosh(c/a)}.
\end{equation}
A particularly attractive feature of the symmetrized Fermi function is that---unlike the conventional 2pF 
function---its form factor can be computed analytically\,\cite{Sprung:1997}. Consequently, all spatial 
moments of the distribution may be computed exactly. In particular, at low momentum transfers, the 
form factor admits the Taylor expansion
\begin{equation}
 \FF{SF}(q) = 1 - \frac{q^{2}}{3!}R^{2} + \frac{q^{4}}{5!}R^{4} + \ldots
 \label{LowqFs}
\end{equation}
where the first two moments are given by\,\cite{Piekarewicz:2016vbn}
\begin{subequations}
\begin{align}
 R^{2}  & \equiv \langle r^{2} \rangle  = \frac{3}{5}c^{2} + 
  \frac{7}{5}(\pi a)^{2} \,,\\
 R^{\,4} & \equiv \langle r^{4} \rangle  = \frac{3}{7}c^{4} + 
  \frac{18}{7}(\pi a)^{2}c^{2} + \frac{31}{7}(\pi a)^{4}  \,.
\end{align}
 \label{SFMoments}
\end{subequations}
This expansion highlights the intimate connection between the low-q behavior of the form factor and the 
spatial moments of the underlying density distribution, with the leading correction determined by the 
mean-square radius. Thus, measurements of the form factor at sufficiently small momentum transfers 
provide direct access to the rms radius of the nucleus.

\begin{center}
\begin{figure}[h]
\centering
\bigskip
\includegraphics[width=0.49\textwidth,height=5.5cm]{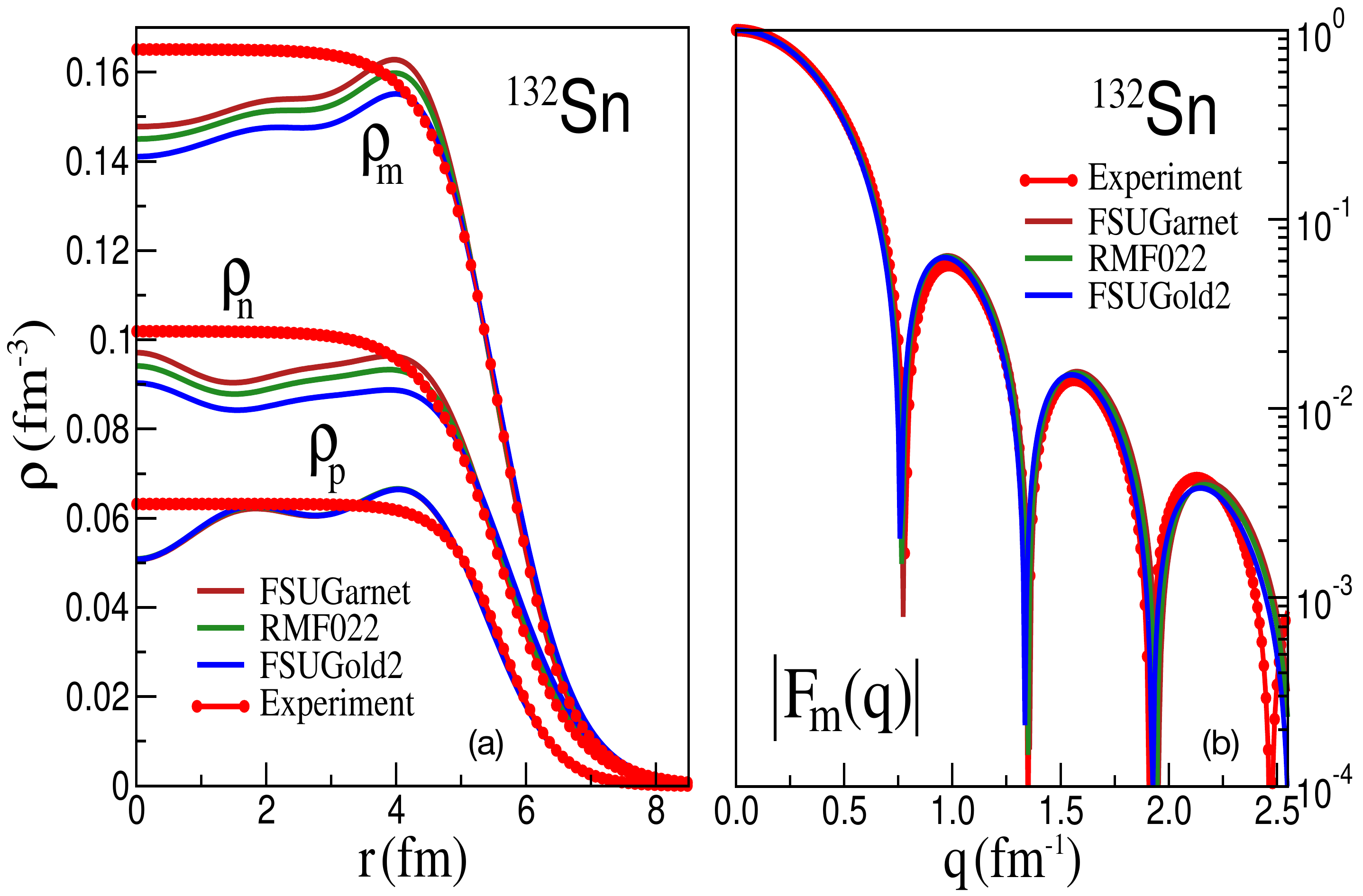}
\caption{(a) Proton, neutron, and matter densities of ${}^{132}$Sn predicted by the three covariant 
energy density functionals used in this work, together with densities extracted from experiment using 
two-parameter Fermi distributions. (b) Corresponding matter form factors. Because protons probe
primarily the nuclear surface, model differences appear only at large momentum transfer, where the 
form factor has already fallen by more than two orders of magnitude.}
\label{Fig2}
\end{figure}
\end{center}

Now displayed in Fig.\ref{Fig2}(a) are proton, neutron, and matter densities as predicted  by the three
covariant energy density functionals listed in Table\,\ref{Table1}. Although agreement on charge radii 
does not guarantee consistency across the entire spatial density, we observe that all three models 
predict nearly identical proton distributions.
This consistency, however, does not extend to the neutron or matter densities. The long-standing 
hope has been that experimental measurements could resolve these differences and thereby constrain 
the density dependence of the symmetry energy. Yet experimental two-parameter Fermi (2pF) distributions 
are---by construction---flat in the nuclear interior and therefore incapable of resolving the detailed internal 
structure of the nucleus. This reflects a generic feature of elastic scattering with strongly interacting probes, 
which are primarily sensitive to the nuclear surface.

The matter form factor, obtained from the Fourier transform of the spatial densities shown in Fig.\ref{Fig2}(b), 
reinforces this picture. Given the surface dominance of the probe, the matter form factor is largely insensitive 
to any model dependence. Differences between the models and the experimental 2pF form factor do not 
emerge until the fourth diffraction maximum, by which point the form factor has already decreased by more 
than two orders of magnitude. Although model differences in the neutron skin thickness are predicted to be 
as large as 60\%, the corresponding differences in the matter radius remain below 2\%. Because the 
characteristic diffractive oscillations of the elastic cross section are governed primarily by the matter radius, 
the large spread in neutron skin predictions among the various models becomes greatly 
diluted\,\cite{Piekarewicz:2005iu}. Listed in Table\,\ref{Table2} are the two-parameter symmetrized Fermi 
distributions for the proton and neutron densities of ${}^{132}$Sn as predicted by the three covariant energy 
density functionals considered in this work, together with those extracted from experiment; see Table 3 of 
Ref.\cite{Hijikata:2026xwa}. A notable difference between the theoretical predictions and the experimental 
extraction is that, in all the theoretical models, the neutron half-density radius is consistently larger than the 
corresponding proton radius.

\begin{center}
\begin{table}[h]
\begin{tabular}{|l|c|c|c||c|c|c|}
\hline\rule{0pt}{2.5ex}   
\!\!Model   &  $c_{p}$  &  $a_{p}$  & $R_{\rm p}$ 
                 &  $c_{n}$  &  $a_{n}$  & $R_{\rm n}$ \\
\hline
\hline
FSUGarnet &  5.637  &  0.412  & 4.627  &  5.709 &  0.536 & 4.850 \\
RMF022     &  5.631 &  0.412  & 4.623  &  5.771 &  0.542 & 4.903 \\
FSUGold2  &  5.630  &  0.415  & 4.626  &  5.867 &  0.545 & 4.976 \\
\hline
Experiment &  5.632   & 0.430  & 4.646 & 5.581 & 0.576 & 4.824 \\
\hline
\end{tabular}
\caption{Two‑parameter symmetrized Fermi distributions for the proton and neutron 
              densities---and the corresponding radii---of ${}^{132}$Sn predicted by the 
              three covariant energy density functionals employed in this work. For the
              experimentally extracted parameters, we list the central values reported in
              Table 3 of Ref.\cite{Hijikata:2026xwa}. All quantities are given in fm.}
\label{Table2}
\end{table}
\end{center}

As we continue to explore the era of rare-isotope beams, it is therefore important to determine whether 
hadronic probes---despite their surface sensitivity---can be refined to provide meaningful constraints on 
the neutron skins of exotic nuclei with large neutron excess. This question is especially relevant given that 
the model-independence of parity-violating electron scattering is experimentally unattainable for unstable
nuclei. Nevertheless, the PREX and CREX measurements provide valuable anchors for guiding new
hadronic experiments and constraining theoretical models.

\begin{center}
\begin{figure}[h]
\centering
\bigskip
\includegraphics[width=0.4\textwidth]{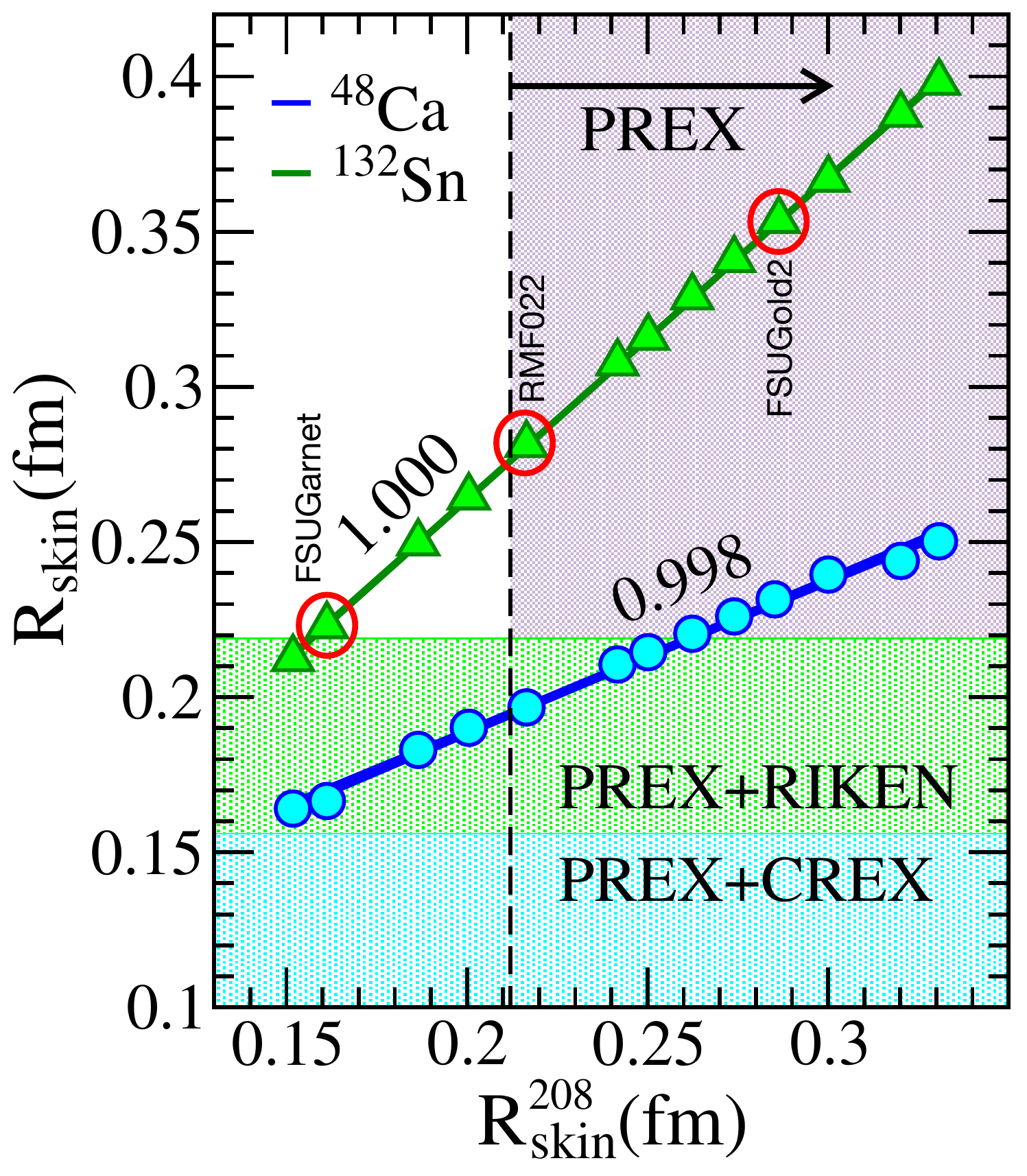}
\caption{Predicted correlation between the neutron skin thickness of $^{208}$Pb, $^{132}$Sn, and 
$^{48}$Ca from a set of covariant energy density functionals. The PREX, CREX, and RIKEN 
constraints are also shown. Indicated with red circles are the predictions from FSUGarnet, RMF022, 
and FSUGold2. The tight correlation illustrates the challenge of reconciling the neutron skins of these 
three nuclei within a single theoretical description.}
\label{Fig3}
\end{figure}
\end{center}

The challenge of extracting the neutron skin from surface-dominated hadronic probes becomes even more 
pressing when considering the current ``CREX--PREX dilemma". As noted earlier, the challenge is not to
reproduce the charge and matter radius of $^{132}$Sn, but rather to do so in the context of the CREX--PREX 
tension. Whereas FSUGold2 is distinctly ``PREX-like" in its prediction of a stiff symmetry energy and a thick 
neutron skin, FSUGarnet is ``CREX-like," favoring a significantly softer symmetry energy and correspondingly 
thinner neutron skin---highlighting a potential deficiency in the current generation of energy density functionals. 
The inclusion of the RMF022 functional serves as an intermediate case between the soft FSUGarnet and stiff 
FSUGold2. 

As a neutron-rich, doubly magic nucleus, $^{132}$Sn offers a unique bridge: if proton scattering in inverse 
kinematics can provide a reliable skin thickness for this isotope, it may clarify the PREX-CREX discrepancy 
and offer a path forward in the determination of the equation of state. The hope is that a single functional can 
consistently describe the evolution of the neutron skin across the nuclear chart. The problem with this scenario 
is that the neutron skin thickness of $^{208}$Pb, $^{132}$Sn, and $^{48}$Ca are highly correlated, so it is 
challenging to reproduce all with current  energy density functionals. In Fig.\ref{Fig3} we display such a correlation 
using 13 covariant energy density functionals of the kind used in this work. Whereas the CREX-like FSUGarnet 
functional is almost consistent with the CREX and RIKEN results, its predictions for the neutron skin thickness 
of $^{208}$Pb is inconsistent with PREX. Conversely, FSUGold2 reproduces the PREX result but grossly 
overestimates the neutron skin thickness reported by both CREX and RIKEN. We note that the strong correlation 
displayed in Fig.\ref{Fig3} extends to other set of functionals---both of relativistic and non-relativistic nature---especially 
in the case of $^{132}$Sn; see Fig.2 in Ref.\,\cite{Piekarewicz:2012pp}.

\begin{center}
\begin{figure}[h]
\centering
\bigskip
\includegraphics[width=0.48\textwidth]{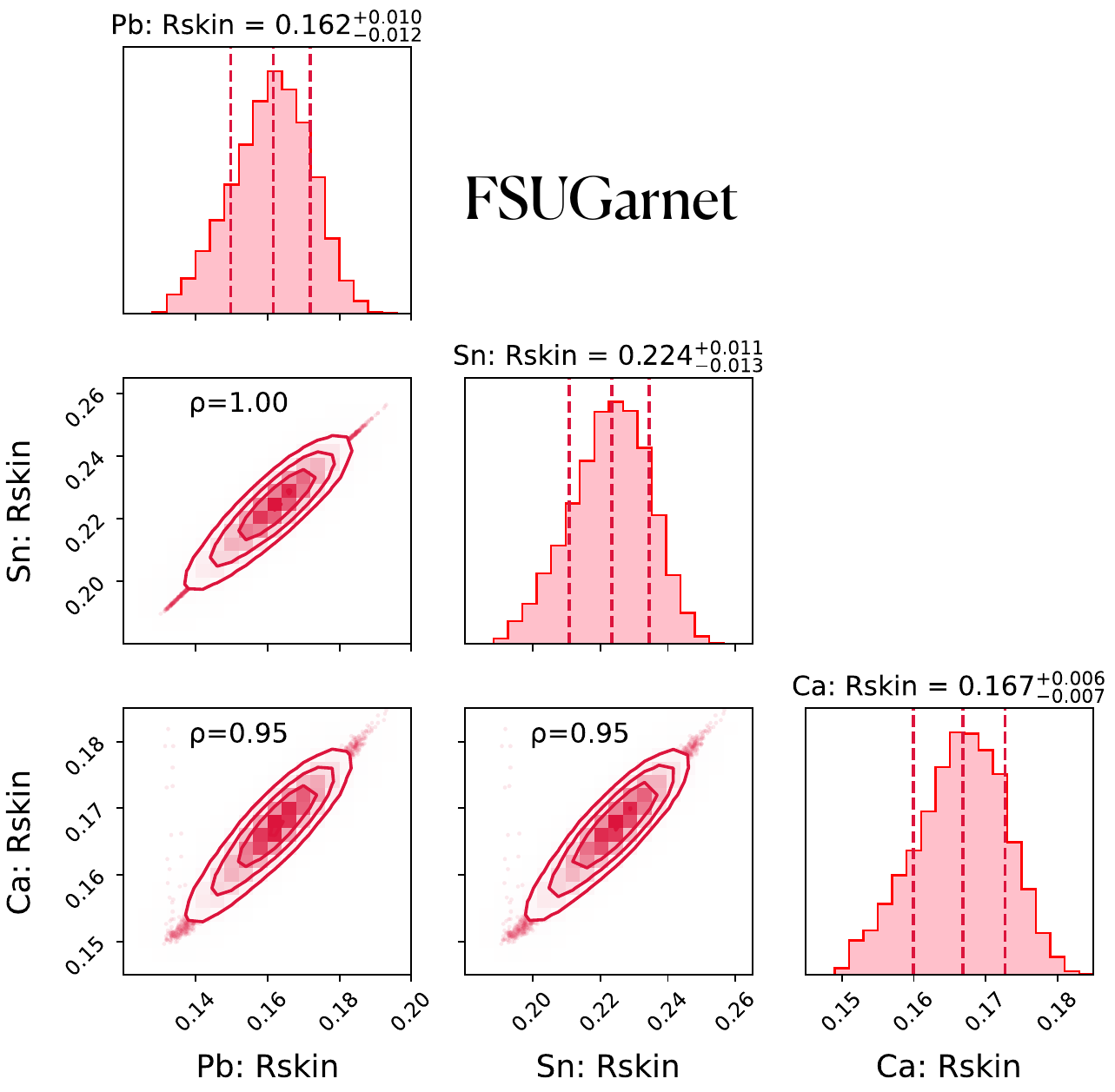}
\includegraphics[width=0.48\textwidth]{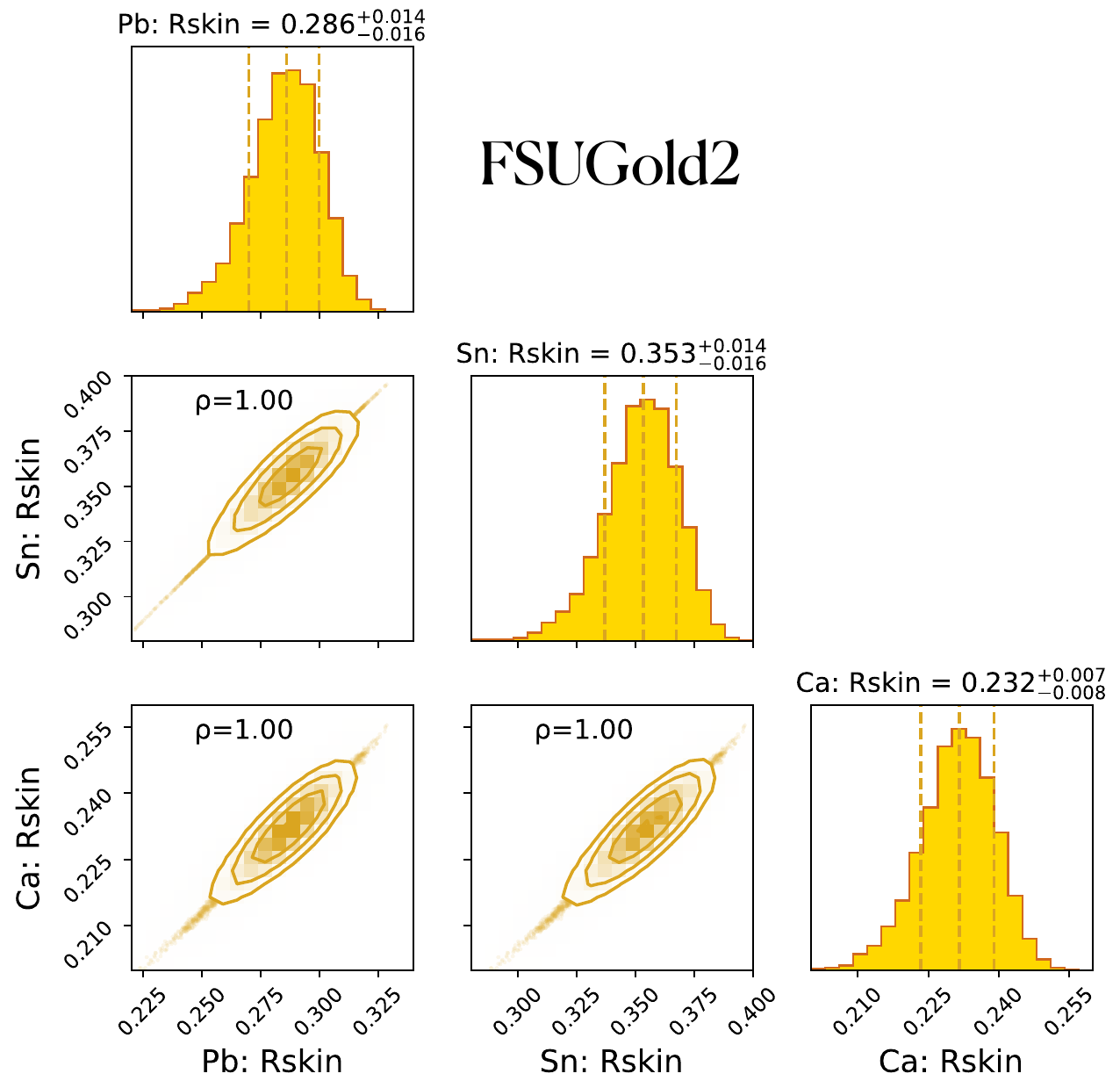}
\caption{Statistical correlations between neutron skin thicknesses obtained from the covariance analysis of the 
              FSUGarnet and FSUGold2 energy density functionals. The neutron skins of $^{208}$Pb, $^{132}$Sn, 
              and $^{48}$Ca exhibit nearly perfect correlations, underscoring the difficulty of simultaneously reproducing 
              all experimental constraints within a single model.}
\label{Fig4}
\end{figure}
\end{center}

Whereas Fig.\ref{Fig3} provides insights into the systematic uncertainties among models that successfully
reproduce ground-state observables, Fig.\ref{Fig4} quantifies the statistical uncertainties within each individual 
model. Displayed in Fig.\ref{Fig4} are the correlations between the neutron skin thickness of ${}^{208}$Pb, 
${}^{132}$Sn, and ${}^{48}$Ca as predicted by the covariance analysis of both FSUGarnet and FSUGold2. 
The figure highlights the robust nature of these correlations: within a given theoretical framework, the predictions 
for these three doubly magic nuclei display---as in the case of Fig.\ref{Fig3}---nearly perfect correlation coefficients.
We observe that while the CREX-like FSUGarnet aligns with the thinner skins reported by CREX and the 
RIKEN measurement, its prediction for $^{208}$Pb falls significantly short of the PREX result. Conversely, 
FSUGold2 accurately captures the thick neutron skin of $^{208}$Pb but overestimates the neutron skins of 
both ${}^{48}$Ca and ${}^{132}$Sn by a fairly wide margin.

In summary, the combined constraints from ${}^{48}$Ca, ${}^{132}$Sn, and ${}^{208}$Pb define a particularly stringent 
test for modern nuclear theory. Like CREX, the recent measurement of the matter radius of ${}^{132}$Sn appears 
to favor a relatively soft symmetry energy, thereby sharpening the tension with the PREX result for ${}^{208}$Pb. 
Reconciling the neutron skins of these three doubly magic nuclei within a unified theoretical framework thus remains 
an outstanding challenge that signals the need for an improved description of the isovector sector of nuclear energy 
density functionals and possibly of the underlying nuclear forces. In this regard, an independent confirmation of the 
PREX result—such as that anticipated from the proposed Mainz Radius EXperiment (MREX) at the MESA facility---will 
be critically important\,\cite{Schlimme:2024eky}.
\vfill

\bibliography{./main.bbl}

\begin{acknowledgments}\vspace{-10pt}
The author is grateful to Profs. K. Kemper and M. Spieker for useful discussions. This material is based 
upon work supported by the U.S. Department of Energy Office of Science, Office of Nuclear Physics under 
Award Numbers DE-FG02-92ER40750 and DE-SC0009883.
\end{acknowledgments}

\end{document}